%% file: paper.tex
\documentclass{article} 
\usepackage{jcappub}

\input{math_commands.tex}

\usepackage{hyperref}
\usepackage{url}
\usepackage{graphicx}
\usepackage{caption}
\usepackage{subcaption}
\usepackage[dvipsnames]{xcolor}
\usepackage{aas_macros}
\usepackage{array}
\newcolumntype{P}[1]{>{\centering\arraybackslash}p{#1}}

\usepackage{algorithm}
\usepackage[noend]{algpseudocode}
\algdef{SE}[VARIABLES]{Variables}{EndVariables}
   {\algorithmicvariables}
   {\algorithmicend\ \algorithmicvariables}
\algnewcommand{\algorithmicvariables}{\textbf{global variables}}
\floatname{algorithm}{Procedure}

\newcommand{\bs}{\mathbf{s}}
\newcommand{\bq}{\mathbf{q}}
\newcommand{\bp}{\mathbf{p}}
\newcommand{\bx}{\mathbf{x}}

\newcommand{\bk}{\mathbf{k}}
\newcommand{\by}{\mathbf{y}}
\newcommand{\bz}{\mathbf{z}}
\newcommand{\bn}{\mathbf{n}}

\newcommand{\bnu}{\boldsymbol{\nu}}
\newcommand{\bmu}{\boldsymbol{\mu}}
\newcommand{\bsigma}{\boldsymbol{\sigma}}
\newcommand{\bbE}{\mathbb{E}}

\author[a,b]{Chirag Modi,}
\author[a,b]{Yin Li,}
\author[b,c]{David Blei}

\affiliation[a]{Center for Computational Astrophysics, Flatiron Institute, New York}
\affiliation[b]{Center for Computational Mathematics, Flatiron Institute, New York}
\affiliation[c]{Columbia University, New York}

\emailAdd{cmodi@flatironinstitute.org}
\emailAdd{yinli@flatironinstitute.org}
\emailAdd{david.blei@columbia.edu}

\title{Reconstructing the Universe with Variational self-Boosted Sampling}

\abstract{Forward modeling approaches in cosmology have made it possible to reconstruct the initial conditions at the beginning of the Universe from the observed survey data.
However the high dimensionality of the parameter space still poses a challenge to explore the full posterior, with traditional algorithms such as Hamiltonian Monte Carlo (HMC) being computationally inefficient due to generating correlated samples and the performance of variational inference being highly dependent on the choice of divergence (loss) function.
Here we develop a hybrid scheme, called variational self-boosted sampling (VBS)
to mitigate the drawbacks of both these algorithms by 
learning a variational approximation for the proposal distribution of Monte Carlo sampling and combine it with HMC. 
The variational distribution is parameterized as a normalizing flow and learnt with samples generated on the fly, while proposals drawn from it reduce auto-correlation length in MCMC chains. 
Our normalizing flow uses Fourier space convolutions and element-wise operations to scale to high dimensions.
We show that after a short initial warm-up and training phase, 
VBS generates better quality of samples than simple VI approaches and 
reduces the correlation length in the sampling phase by a factor of 10-50 over using only HMC to explore the posterior of initial conditions in 64$^3$ and 128$^3$ dimensional problems, with larger gains for high signal-to-noise data observations. 
}

\begin{document}
\maketitle
\flushbottom

\section{Introduction}
 
Forward modeling approaches for cosmological analysis is one of the most promising ways to fulfill the scientific potential of the current and upcoming cosmological surveys such as DESI \cite{DESI}, LSST \cite{LSST}, Euclid \cite{EUCLID18} and others.
In this approach, we simulate the field level cosmological survey data 
such as galaxies, starting all the way from the cosmological parameters and the initial conditions at the beginning of the Universe. 
This allows one to make model comparison at the field level, thus maximizing the amount of information that can be extracted from these surveys as one no longer relies on any compressed statistics of the survey data \citep{Seljak17}.
However, at the same time, it makes inference challenging due to the much 
larger dimensionality of the parameter space since to infer cosmological parameters, we now need to marginalize over the density field at all points in the Universe.

Recent works have investigated a number of ways to infer the Gaussian initial conditions of the Universe in a Bayesian framework. 
The common theme amongst them is to begin by constructing a posterior of the initial conditions by combining the Gaussian prior on them with the data likelihood at field level.
To make inference tractable in high dimensions, these then use differentiable simulations \citep{flowpm, madlens, borg} which allow one to analytically estimate response of the observed data with respect to these underlying parameters.

The simplest inference is to reconstruct a maximum-a-posterior (MAP) estimate by maximizing this posterior with traditional gradient based optimization algorithms \citep{Seljak17, Modi18, Modi19}
or by using learnt optimization \citep{cosmicrim}.
While being the fastest way, this provides only a point estimate for the initial conditions and one can estimate uncertainties by making a Laplace approximation \citep{Seljak17, Horowitz18, muse}.

A more robust but expensive approach to infer the posterior is to use Markov Chain Monte Carlo (MCMC) methods \citep{borg, Wang14, Kitaura14}, particularly Hamiltonian Monte Carlo (HMC) \citep{Duane87, Neal11, Betancourt17}. 
HMC generates samples from the posterior by simulating a Markov chain following Hamiltonian dynamics for multiple steps and thus minimizes the random walk diffusive behavior between successive proposals by taking longer jumps.
Despite this, successive samples generated by HMC are still correlated and these correlation lengths can be hundreds of samples long.
Hence overall, the cost of this approach is prohibitively expensive for scaling up to the future cosmological surveys because one requires at least on the order of hundred independent posterior samples to ensure that the first two moments (mean and variance) are estimated correctly. 

In this work, we take a step towards reducing this computational cost of sampling approaches by learning a proposal distribution and combining it with MCMC algorithms. 
We parameterize this proposal distribution as a normalizing flow \citep{Kobyzev20} which is trained using samples from the MCMC chain itself. 
Then the goal is that the independent samples generated from this proposal kernel either lie in the target distribution directly, or can propagate to the same with short Markov chains. 
In statistics literature, similar approaches have been proposed to speed up HMC by improving or learning the geometry of the posterior distribution with a transport map \citep{Hoffman19, Naesseth20}.
However our scheme most closely resembles the approach proposed in \cite{Gabrie21}
which uses the variational distribution as a global sampler to facilitate mode jumping in multi-modal distributions.

Another perspective on our approach is lent by variational inference itself \citep{Blei17}.
The learnt proposal distribution for MCMC can instead be viewed as a variational approximation to the target distribution with the associated short Markov chains serving a corrections to the learnt approximation.
Alternatively, turning things around yet again, we can also view this as first running MCMC to generate samples from the true distribution which can then be used for learning the variational parameters by minimizing a more constraining forward (inclusive) Kullback-Leibler (KL) divergence \citep{KLdivergence}, and then using this learnt distribution to generate uncorrelated proposals and speed up MCMC. 

The challenge then remains to be able to learn this high dimensional variational distribution with normalizing flows (NF). 
Scaling to three dimensional data and distributions with millions of parameters is still not feasible with commonly used architectures of normalizing flows.
In this work we build upon a recently developed architecture that exploits rotational and translational symmetries for cosmological fields by performing convolutions in the dual Fourier space \citep{TRENF} instead of the usual grid (pixel) space. We show that with suitable modifications to the input base distribution of NF, our model is flexible enough to learn a good proposal kernel of HMC to explore the posterior distribution of the initial conditions. 

The paper is organized as follows. 
We begin by setting up our cosmological inference problem formally in section \ref{sec:setup} with the relevant details and notation. 
Then we briefly review the Hamiltonian Monte Carlo (HMC) and Variational Inference (VI) algorithms in section \ref{sec:inference}. 
We also discuss their performance on our example problem to set up a benchmark. 
Then we present our 
hybrid sampling scheme with in \ref{sec:main}.
For coherence, we have moved the detailed discussion of our normalizing flow into the appendix \ref{app:nflows}.
Finally we present the results in section \ref{sec:results} and conclude in section \ref{sec:conclusion}.

\section{Setup}
\label{sec:setup}

We are interested in forward modeling approach for cosmological inference wherein we begin with the cosmological parameters and the initial dark matter density field, build a forward model for the data, and use this to infer the posterior of the parameters as well as the initial conditions. 
In this first work, we use the final evolved dark matter density as the data.
Furthermore, we focus only on inferring the posterior of the initial density field while keeping the cosmological parameters fixed to their true value since this is the challenging high-dimensional part of the problem. 

\subsection{Data and Likelihood model}
Our data ($\by_0$) is the dark matter density field on a cubic $N^{3}$ grid\footnote{Unless specified otherwise, all the bold-face symbols such as $\bx,\, \by,\, \bs,\, \bz$ are $N^3$ vector corresponding to the cubic simulation grid. We refer to these as fields} where N is the number of grid points or pixels along each side of the cube.
For the toy problem in this work, we consider small grids of N=64 and 128 while analyzing future cosmological surveys will require scaling up to N=256 or 512.
We choose the boxsize between L=500 Mpc/h and 1000 Mpc/h so as to keep the resolution of our toy model simulations same as future realistic problems.

The data has been generated from some unknown initial conditions, i.e. initial dark matter density field ($\bs$) which is evolved under gravity with a realistic forward model ($f$) to simulate a final dark matter field ($\by$) and then corrupted with a noise model ($\bn$). 
The simple forward model we use is the particle displacement predicted by the first order Lagrangian Perturbation theory (Zeldovich Approximation, ZA). 
We use ZA purely for computational reasons and expect our conclusions to qualitatively remain the same for more realistic forward models such as FastPM\citep{Feng2016} or COLA\citep{Tassev13}.
We take our data noise to be Gaussian with known noise variance ($\bn \sim \mathcal{N}(\mathbf{0}, \bsigma)$) corresponding to the shot-noise of the dark matter particles. 
This allows us to compute the exact likelihood for our toy data. 

The parameters to be inferred are the phases of the Gaussian field ($\bz \sim \mathcal{N}(\mathbf{0}, \mathbf{I})$) corresponding to this unknown initial dark matter density field ($\bs$).
There is a deterministic relationship between the cosmology parameters ($\Lambda$), the phases and the initial density field, $\bs = g(\bz, \Lambda)$. 
Then for our toy problem, it is identical whether we infer $\bs$ or $\bz$ since we keep $\Lambda$ fixed to their true value, 
but we choose to work with $\bz$ here in the anticipation for follow up works
when $\Lambda$ needs to be inferred simultaneously with $\bz$.

Then given the parameters $\bz$, the likelihood model, prior, and the posterior distribution is 
\begin{align}
    \pi(\by_0 |\bz) &= \mathcal{N}(\by = f(\bz), \bsigma) \tag {Gaussian likelihood} \\
    \pi(\bz) &= \mathcal{N}(\mathbf{0}, \mathbf{1}) \tag {Gaussian prior} \\
    \pi(\bz , \by_0) &= \pi(\by_0 | \bz) \pi(\bz) \tag {Unnormalized Posterior}
\end{align}

\begin{figure}[h]
\begin{center}
\resizebox{\columnwidth}{!}{\includegraphics{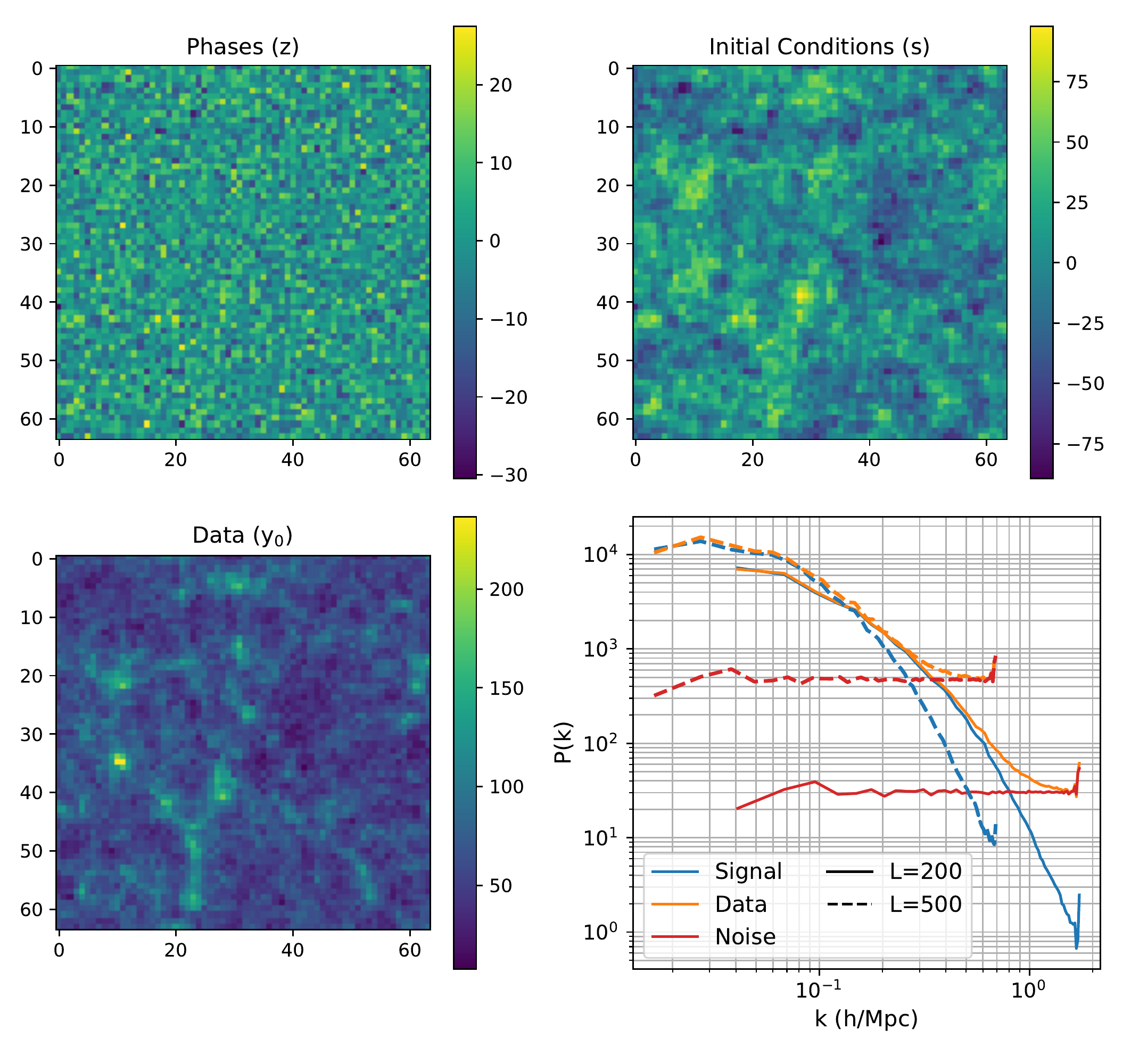}}
\end{center}
\caption{An example of the toy model: first three panel show different fields involved in our simulation for a L-200 Mpc/h and N=64 grid. The fields are projected (summed) along the z-axis. The last panel shows the corresponding power spectrum for two different box sizes- L=200 Mpc/h and 500 Mpc/h which correspond to different signal-to-noise ratio.
}
\label{fig:data}
\end{figure}

In Figure \ref{fig:data}, we show the different component fields of our problem for N=64 grid: the phases ($\bz$), the initial conditions $(\bs)$, and the data ($\by_0$, final dark matter field with noise).
In the last panel, we also show the power spectra of the data signal and noise for two different box sizes which correspond to different signal to noise ratios.


\subsection{Validating the posterior}
Before discussing different approaches to inference, we briefly lay out the metrics that will be used to validate the posterior distribution inferred by them. 
We will do so by looking at the samples $\bz_i$ (and their derived properties as described below) generated from these distributions. 

\subsubsection{Distribution of summary statistics}
We are trying to infer the posterior of the phases of the initial conditions.
This is a high dimensional density distribution, and they are notoriously hard to compare quantitatively.
Thus we seek a low dimensional mapping to compare posterior samples from different algorithms. 
Our data model obeys rotational and translational invariance and hence the power spectrum of density fields provides a natural low dimensional candidate for this mapping.
The power spectrum $(P_a)$ of any field $a$ measures the clustering of the overdensity field $\delta_a$ at different scales $\bk$ and is defined as
$$ \langle {\tilde {\delta_a}}(\mathbf {k} ){\tilde {\delta_a}}^{*}(\mathbf {k} ')\rangle =(2\pi )^{3}P_a(k)\delta ^{3}_D(\mathbf {k} -\mathbf {k} ')
$$
where $k$ is the magnitude of the scale and $\delta^3_D$ is the 3-D Dirac-delta function. 
We will measure the quality of the posterior distributions with the following two derived quantities of the power spectra of the posterior samples: 
\begin{enumerate}
    \item Cross correlation ($r_c$) of the samples from the posterior with the true initial conditions defined as
    $$r_c = \frac{P_{ab}(k)}{\sqrt{P_a(k) P_b(k)}}$$
    for any two fields $a$ and $b$ and where $P_{ab}$ is their cross-spectra. 
    Cross correlation between two fields effectively compares the phases of the fields i.e. if the features such as peaks and voids of the density field are physically at the same location on different scales.
    We expect $r_c$ to be consistent with unity on large scales which are signal dominated and then drop to zero on the scales where our posterior is prior dominated. 
    \item Transfer Function ($t_f$) of the samples from the posterior with the true initial conditions defined as
    $$t_f = {\sqrt{P_b(k) /P_a(k)}} $$
    for any two fields $a$ and $b$.
    Transfer function compares the amplitude of clustering at different scales. Since we use the same cosmology for data generation and inference, $t_f$ of samples from the correct posterior should be consistent with unity on all scales. 
    
\end{enumerate}

We will show these summary statistics for the posterior samples which are the phases of the initial dark matter density field $\bz$.
Thus in our discussion, the two fields in comparisons are
\begin{align}
    a &:= \bz_0 \qquad \textrm{Phases of the true initial dark matter field corresponding to data $\by_0$} \nonumber \\
    b &:= \bz_i \qquad \textrm{Posterior sample $\bz_i$} \nonumber 
\end{align}
Note that when both $r_c$ and $t_f$ are unity, the two fields being compared are identical.

\subsubsection{Auto-correlation length}

Monte Carlo algorithms explore the posterior by generating samples from it instead of optimizing (learning) a parametric form of them. 
In this case it is important to have generated enough independent samples such that we are confident to have explored both, the bulk and the tails of the posterior adequately. Thus the efficacy of such algorithms is measured with auto-correlation length which is the effective length (number of samples) between two successive independent samples. 

As discussed above, due to the high dimensional nature of our problem, we will again work with low-dimensional summary statistic for quantitiative comparisons. 
Hence we compare the efficacy of algorithms by estimating the auto-correlation length for power spectrum of the posterior samples. 
Specifically, for every chain, we measure the power spectrum $P_i(k)$ for each sample $\bz_i$
and then estimate the correlation length for each mode $k_j$ as
\begin{equation}
    \rho_j(t)  = \frac{1}{n} \sum_{i=t+1}^n (P_i(k_j) - \bar{P}(k_j)) (P_{i-t}(k_j) - \bar{P}(k_j))
    \label{eq:ac_lenght}
\end{equation}
where $\bar{P}(k_j)$ is the mean power in the mode $k_j$ across all samples of that chain 
and $n$ is the total number of samples. 
Then the auto-correlation length ($a_c$) is defined as the scale where $\rho_j(a_c) \leq 0.1$.
We want the auto-correlation $a_c$ as small as possible since it implies more independent samples for the same computational cost.

\section{Posterior of initial conditions}
\label{sec:inference}
We begin by briefly reviewing the two most widely used approaches for posterior inference, which will also form the building blocks of our hybrid sampling presented in the next section. These are- 
i) Hamiltonian Monte Carlo (HMC) which generates samples from the posterior directly and
ii) Variational Inference (VI) which learns a parametric form of the posterior distribution.
For each approach, we will also discuss their merits, drawbacks, and evaluate their performance for our particular problem to infer the posterior of the initial conditions. 

\subsection{Hamiltonian Monte Carlo (HMC)}
\label{sec:hmc}

\begin{algorithm}[t]
\caption{Single step of Hamiltonian Monte Carlo Sampling}
\label{alg:hmc}
\begin{algorithmic}[1]
\Require{ {current position $z_0$;
            target probability density $\pi$;
            step-size $\epsilon$;
            number of leapfrog steps $L$;
            Hamiltonian $H$;
            Mass matrix for momentum $M$
    }
}
\State{$q_0 \leftarrow z_0 $} \Comment{Assign current sample as the initial position}
\State{$p_0 \sim \mathcal{N}(0, 1)$} \Comment{Sample random momentum of same shape as $q_0$}
\State{$i = 0$}
\While{$i \leq L$} \Comment{Integrate Hamiltonian equations for $L$ leapfrog steps}
    \State{$q_{i+1},\ p_{i+1} \leftarrow$ \Call{Leapfrog}{$q_i,\ p_i,\ \pi, \epsilon$}} 
    \State{$i \leftarrow i+1 $}
\EndWhile
\State{$H_0 \leftarrow$ \Call{H}{$q_0,\ p_0, \pi, M$}}  \Comment{Estimate Hamiltonian using Eq.~\ref{eq:H}}
\State{$H_L \leftarrow$ \Call{H}{$q_L,\ p_L, \pi, M$}} 
\State{$\alpha \leftarrow \exp(H_0 - H_L)$}             \Comment{Acceptance probability to maintain DB Eq.~\ref{eq:DB}}
\If{$\mathrm{Uniform}(0, 1) \geq \alpha$}
    \State{$z_1 \leftarrow q_0$}
\Else
    \State{$z_1 \leftarrow q_L$}
\EndIf
\Ensure{$z_1$}
\end{algorithmic}
\end{algorithm}

HMC is a widely used approach to generate samples from distributions in high dimensions. 
It begins by reinterpreting the parameters of interest as a position vector $\bq \in R^d$
with the associated potential energy function $U(\bq) = -\log \pi(\bq)$ where  $\pi(\bq)$ is the target distribution (in this case, the unnormalized posterior),
and introducing an auxiliary momentum vector $\bp \in R^d$ which contributes a kinetic energy term $K(\bp) = \frac{1}{2}\bp^T M^{-1}\bp$, where $M$ is a symmetric positive definite mass matrix. 
Generally the mass matrix is taken to be the identity matrix, ${M}$ = ${I}$.
With these, one can construct the Hamiltonian $H:\mathcal{R}^{2d} \to \mathcal{R}$ as the total energy function for the state $\bx := (\bq,\bp)$,
\begin{align}
H(\bx) \equiv H(\bq,\bp) &= U(\bq) + \frac{1}{2}\bp^T M^{-1} \bp \nonumber \\
&= -\log \pi(\bq) + \frac{1}{2}\bp^T M^{-1} \bp~.
\label{eq:H}    
\end{align}
The goal is to simulate a Markov chain and generate samples from the target distribution.
This is achieved by evolving this physical system with respect to time by following Hamiltonian dynamics.
The Hamiltonian's equations are numerically evolved by integrating the ODE system using leapfrog integrator \citep{Neal11}.
Hence there are two parameters to be tuned- the stepsize of the integration $\epsilon$ and the number of leapfrog steps ($L$) to take before making a proposal $\bq_i$.

Proposals generated at the end of each iteration are accepted or rejected to maintain a detailed balance (DB) condition
which guarantees that the samples are generated from a stationary target distribution.
As per detailed balance, the probability of accepting a proposal $\bx_0 \rightarrow \bx_1$ is
\begin{equation}
\alpha = \mathrm{min}(1, \exp(H(\bx_0) - H (\bx_1))
\label{eq:DB}    
\end{equation}

The complete algorithm for generating proposals is described in Algorithm \ref{alg:hmc}. A more in-depth discussion of HMC can be found in \citep{Duane87, Neal11, Betancourt17}.

\subsubsection{Cost of HMC}

\begin{figure}
\begin{center}
\resizebox{\columnwidth}{!}{\includegraphics{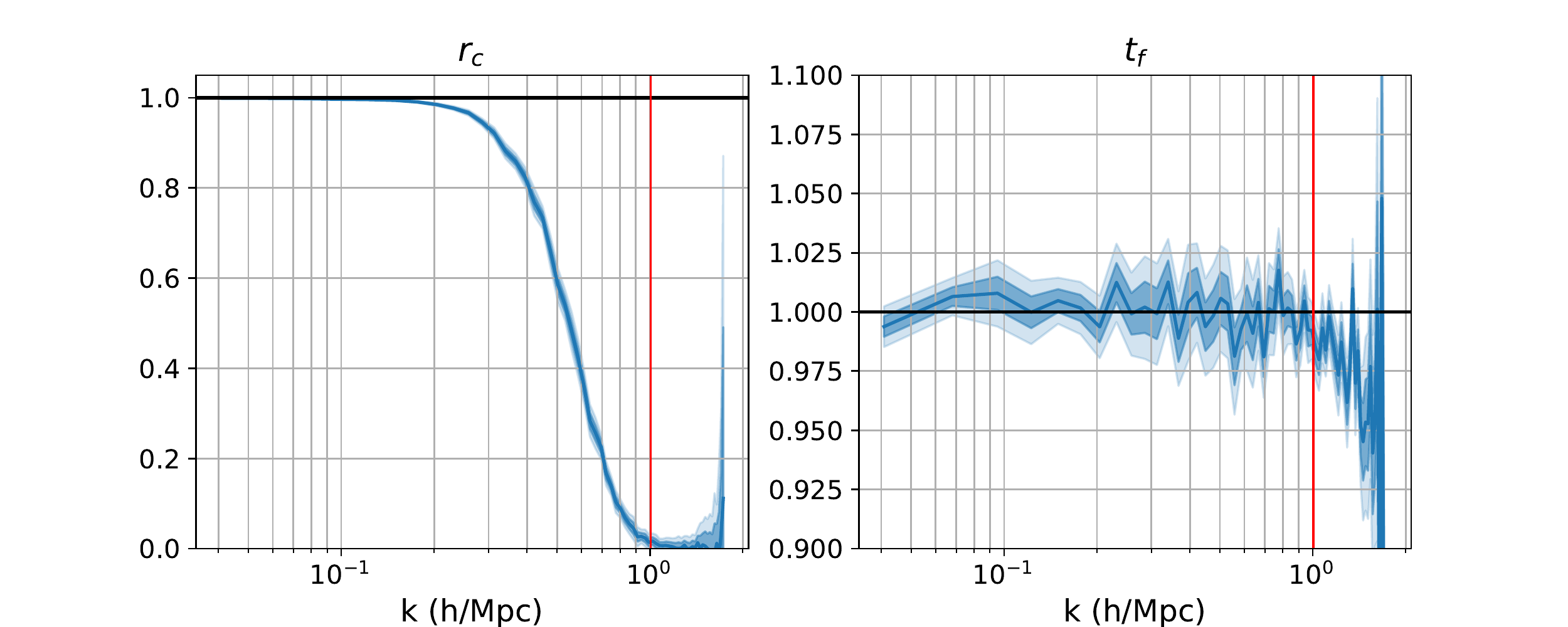}}
\end{center}
\caption{Posterior explored with HMC: We show the distribution of summary statistics for samples from posterior generated with HMC after sub-sampling by a factor of 20. The results are shown for our fiducial configuration of L=200 Mpc/h and N=64 simulation.  The vertical red line represents nyquist frequency. Note both cross correlation and transfer function being 1 implies a perfect inference.}
\label{fig:hmc}
\end{figure}

We set up HMC for our toy problem with the parameters of interest, $\bz$ corresponding to the position vector in HMC $\bq$.
We fit $\eps$ by dual averaging scheme \citep{NUTS} and based on some preliminary experiments, we randomly choose the number of leapfrog steps $L$ uniformly between 25 and 50 for every proposal. 
We run 4 independent chains for robustness, generating 5000 samples in each chain and then thinning them by a factor of 20.
Figure \ref{fig:hmc} shows the summary statistics for the posterior samples
generated by HMC for the configuration with N=64 and L=200 Mpc/h simulation. 
Both the cross correlation and transfer function follow the expected behavior---the cross correlation is one on large scales and falls to zero on small scales while the transfer function is unity across all scales upto the nyquist frequency.

\begin{figure}[h]
\begin{center}
\hspace*{-0.6cm}\includegraphics[scale=0.43]{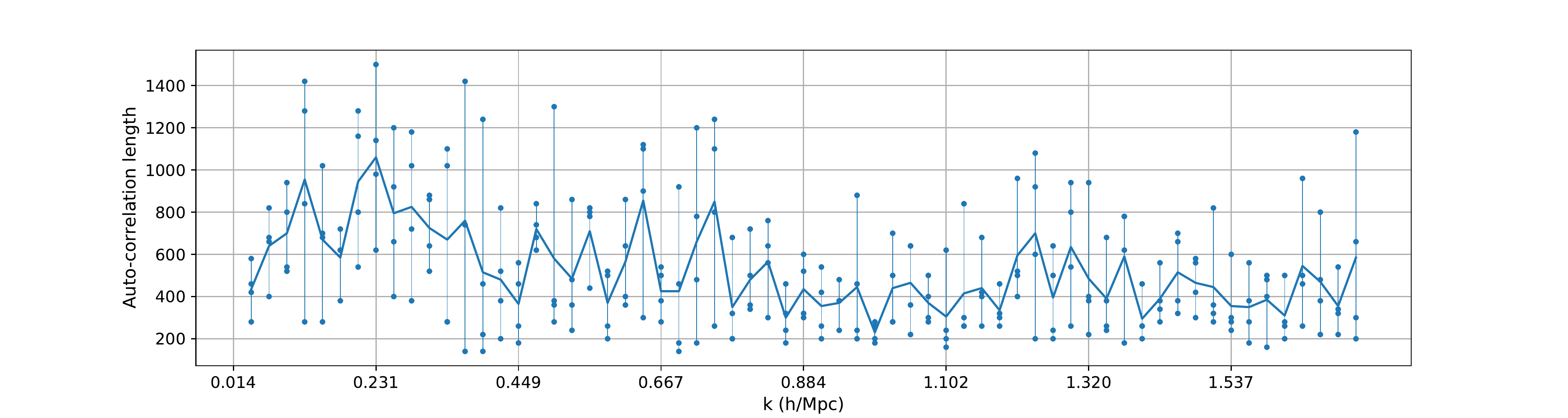}
\end{center}
\caption{Cost of HMC: We show auto-correlation length of HMC samples as estimated for the power in different modes for our fiducial configuration of L=200 Mpc/h and N=64 . 
Different points along the same vertical (k-mode) are four different chains.
%
}
\label{fig:hmc-ac}
\end{figure}

While HMC samples the distribution correctly, it is only true in asymptotic sense and
consecutive samples are still correlated which makes the algorithm computationally expensive.
To gauge it's efficacy, we estimate the auto-correlation length of the chains in terms of their power spectra as described in Eq.~\ref{eq:ac_lenght}.
This is shown in Figure \ref{fig:hmc-ac} as function of different scales. 
The correlation length is larger on the largest scales which are more signal dominated than the noisy small scales. 
On the largest scales, the correlation length reach upto a 1000 samples long.
Taking into account the fact that each sample is generated after taking $L \in [25, 50]$ leapfrog steps,
the cost of a single independent sample on these largest scales can easily be up to $\sim$ 10000 forward simulations. 
This makes HMC prohibitively expensive for scaling to problems of larger size and higher signal-to-noise
since one expects the correlation lengths in these cases to be longer still. 
However due to the algorithm's guaranteed asymptotic correctness, we will use HMC samples as benchmark to compare other posteriors throughout this work.

\subsection{Variational Inference}
\label{sec:vi}
Variational inference \citep{Blei17} takes a different approach from sampling and instead approximates the target distribution $\pi$ with a distribution belonging to a parametric family, $q(\bnu)$.
The parameters $\bnu$ are estimated so as to minimize a divergence between the variational distribution $q(\bnu)$ and the target distribution $\pi$. 
Since this minimization is an optimization, VI is generally much faster than HMC but does not enjoy the guarantees of asymptotic correctness same as HMC. 
In fact, as we will show in this section, the quality of VI depends significantly on the the choice of parametric family and the objective function which is used for this optimization. 

\subsubsection{Normalizing Flow}
In this work, we approximate the posterior of the phases of the initial conditions $\pi(\bz|\by_0)$ with a normalizing flow (NF) \citep{Kobyzev20}.
Normalizing flows consist of a series of invertible, bijective mappings that successively transform (transport) a simple base distribution to a complex distribution with non-trivial correlations. 
Let this transport map be given by $T_{\theta}$ with parameters $\theta$ and the base distribution be given by $q_B$.
Then the variational family of our posterior as parameterized by the normalizing flow is \begin{equation}
    q(\bz; \bnu) = q_B(T_\theta^{-1}(\bz); \bnu_B) |\det \nabla_{\bz} T^{-1}_\theta |
\end{equation}
The set of variational parameters consists of both, the parameters of the base distribution as well as the transport map $\bnu=\{\bnu_B, \theta\}$.

For our base distribution, we choose the mean-field normal i.e. $q(\bz; \bnu_B) := \mathcal{N}(\bmu, \bm\Sigma)$ with a diagonal covariance matrix.
Our transport map is inspired by the fact that the distribution of cosmological fields is rotationally and translationally invariant \citep{TRENF}.
Building upon these invariances, we can model the convolutions as simple products in Fourier space.
This maintains bijectivity and easy Jacobian evaluations which allows us to learn the high dimensional distributions of interest. 
We alternate these Fourier convolutions with other element-wise bijective operations. 
Further details of our normalizing flow are given in the appendix \ref{app:nflows} where we also demonstrate that this variational distribution is flexible enough to learn a good approximation for the target posterior distribution.

\begin{algorithm}[t]
\caption{Backward/Exclusive Variational Inference}
\label{alg:evi}
\begin{algorithmic}[1]
\Require{ {variational family with parameters $\bnu$ $q(\bz; \bnu$);
            likelihood function $\pi(\by_0|\bz)$; 
            prior $\pi(\bz)$; 
            step-size for optimizer $\epsilon$; 
            maximum number of iterations $N$;
            number of samples per iteration $n$
            }}
\State{$i = 0$}
\While{$i \leq N$} 
    \State{$\{\bz_i...\bz_n\} \sim q(\bz;\bnu)$}    \Comment{Generate $n$ samples from variational distribution}
    \State{$\mathrm{ELBO} = \sum_{\bz_i}  \log \pi(\by_0 | \bz_i) + \log \pi(\bz_i) - \log q(\bz_i; \bnu)$}
    \State{$\bnu \leftarrow \bnu - \epsilon \nabla_{\bnu} \mathrm{ELBO}$}  \Comment{Optimization}
\EndWhile
\State{$\bnu^* \leftarrow \bnu$}
\Ensure{$q(\bz; \bnu^*)$}
\end{algorithmic}
\end{algorithm}

\subsubsection{Backward or Exclusive KL Divergence}
\label{sec:bvi}
The other component of variational inference 
is the choice of divergence to be minimized between the variational distribution and the target distribution. 
The most commonly used divergence is Kullback-Leibler (KL) divergence with 
the variational distribution as the reference distribution, In this case its called backward or exclusive KL divergence
 and is defined as
 \begin{align}
    D_{\mathrm{KL}}(q||p) &= \bbE_{q}(\log q - \log p) \nonumber \\
            &= \bbE_{q}(\log q(\bz; \bnu) - \log \pi(\bz| \by_0)) \nonumber \\
            & \approx \sum_{\bz_i \sim q(\bz)} \bigl[ \log q(\bz_i; \bnu) - \log \pi(\bz_i | \by_0)\bigr] \nonumber \\
            & \leq \sum_{\bz_i \sim q(\bz)} \bigl[ \log q(\bz_i; \bnu) - \log \pi(\by_0 | \bz_i) - \log \pi(\bz_i) \bigr]
\label{eq:ekl}
\end{align}
where in the third line we have approximated the expectation with empirical expectation as estimated by the samples $\bz_i \sim q(\bz; \bnu)$ from the variational family.
In the last line, we expand the posterior distribution in terms of the likelihood and the prior while dropping the evidence term which is a negative constant with respect to the variational parameters. 
This is also called the evidence lower bound (ELBO)
\begin{equation}
    \mathrm{ELBO} := \sum_{\bz_i \sim q(\bz)}  \log \pi(\by_0 | \bz_i) + \log \pi(\bz_i) - \log q(\bz_i; \bnu)
    \label{eq:elbo}
\end{equation}

For inferring the posterior, backward VI maximizes the ELBO with respect to the variational parameters.
\begin{equation}
    \bnu^* = \mathrm{arg max}_\nu \mathrm{ELBO}
    \label{eq:ekl*}
\end{equation}
The full algorithm for this is given in Algorithm \ref{alg:evi}

\begin{figure}
\begin{center}
\resizebox{1.0\columnwidth}{!}{\includegraphics{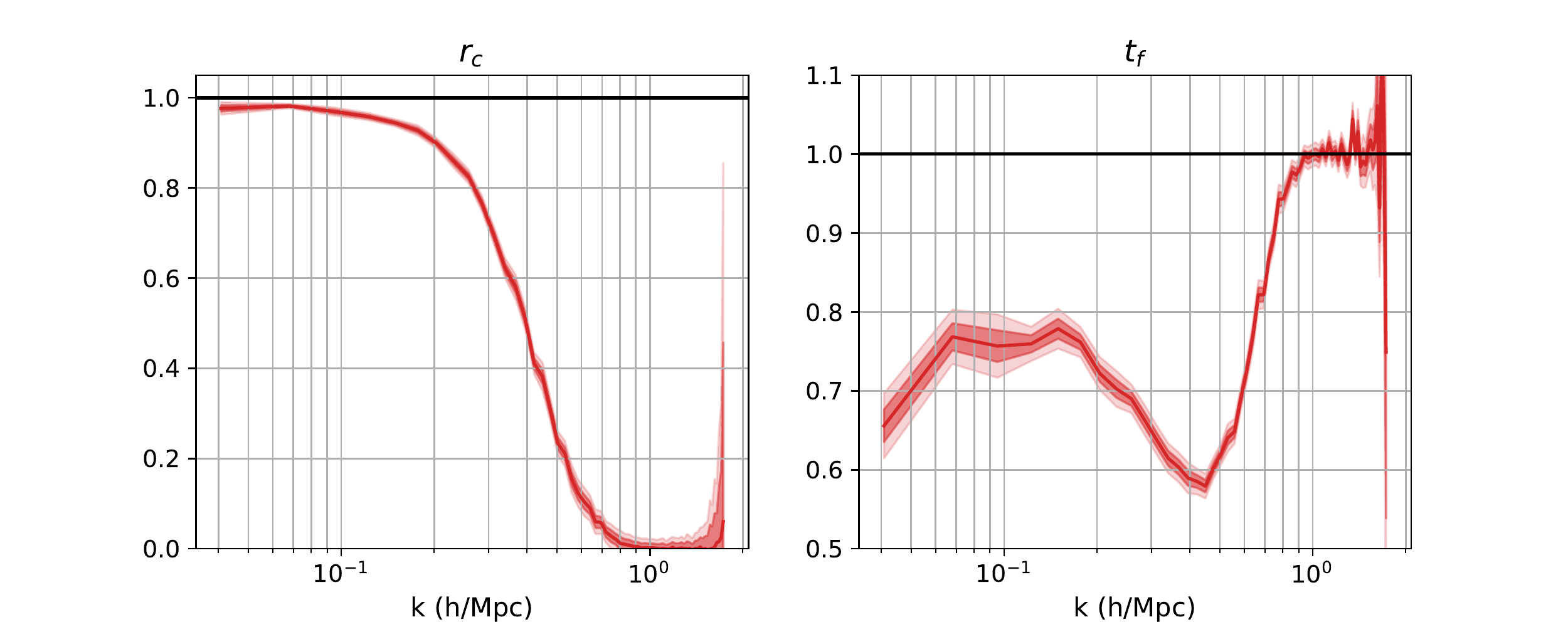}}
\end{center}
\caption{Posterior learnt with backwards VI: We show the cross correlation and transfer function for samples from the variational distribution fit by maximizing ELBO as in Algorithm \ref{alg:evi}.
The result is shown for our fiducial configuration of L=200 Mpc/h and N=64 simulation.
The solid lines and shaded regions show the mean and 1,2$\sigma$ variations of 100 samples generated from the variational distribution after fitting.
}
\label{fig:bbvi}
\end{figure}

We learn the posterior for the same configuration of our toy problem as before, i.e. N=64 and L=200 Mpc/h simulation, with the aforementioned normalizing flow and by minimizing the backward KL loss. 
In Figure \ref{fig:bbvi}, we show the summary statistics for the samples generated by the learnt variational distribution.
Note that while the cross-correlation of the samples is similar to the correct samples generated by HMC, the transfer function is wrong.
There can be two potential reasons for this- 
i) our parametric family i.e. the combination of mean-field Gaussian and the normalizing flow is not flexible enough to model the correct posterior, 
ii) our choice of divergence is not appropriate to learn the correct distribution. 
However as we show in appendix \ref{app:nflows}, our normalizing flow is indeed powerful enough to learn the distribution of samples generated from HMC (atleast at the level of the summary statistics).
This means that in our case, the backward KL loss is not constraining enough to learn the correct variational distribution. 
This is not entirely unexpected since backward KL loss is known to have mode-seeking behavior and does not guarantee coverage of the full target distribution. 

\subsubsection{Forward or Inclusive KL Divergence}
\label{sec:fvi}
An alternative to backward KL divergence is the forward KL divergence
which uses the target distribution as the reference. Then
\begin{align}
    D_{\mathrm{KL}}(p||q) &= \bbE_{p}(\log p - \log q) \\
                &= \bbE_{\pi(\bz | \by_0)} ( \log \pi(\bz | \by_0) - \log q(\bz; \bnu)) \\
                &\approx \sum_{\bz_i \sim \pi(\bz | \by_0)}(\log \pi(\bz | \by_0) - \log q(\bz; \bnu))
\label{eq:ikl}
\end{align}
where we have again approximated the expectation with empirical expectation.
Note that since the samples are generated from the target distribution itself, the first term is independent of the variational parameters. 
Thus minimizing this divergence for variational inference is achieved by maximizing the log-probability of the samples under the variational distribution
\begin{equation}
    \bnu^* = \mathrm{arg max}_\nu \sum_{\bz_i \sim \pi(\bz | \by_0)} \log q(\bz; \bnu)
    \label{eq:ikl*}
\end{equation}

Looking at this equation, we can see the chicken-and-egg problem of the forward KL loss---we need samples $\bz_i$ from the target distribution (e.g., as generated by HMC) to learn the variational distribution, but if we had an easy access to such samples, we would not need to learn a variational distribution in the first place.
Recent works have investigated some ways to get around this, such as with importance weighing the samples generated from the variational distribution \citep{Naesseth20, RWS}.
However we find that none of these approaches work well in our case.
Hence in the next section, we turn back to HMC and VI to combine them in a hybrid approach that benefits from the complementary advantages of both the algorithms. 

\begin{algorithm}[t]
\caption{Variational (self-)Boosted Sampling}
\label{alg:hybrid}
\begin{algorithmic}[1]
\Require{ {Initial sample from the target distribution $\bz_0$;
           variational family $q(\bz; \bnu)$;
            target distribution (posterior) $\pi(\bz |\by_0)$; 
            annealed target distribution for VI $\pi^*(\bz|\by_0)$; 
            step-size for HMC $\epsilon$;
            step-size for training $\epsilon_q$; 
            number of leapfrog steps $L$;
            mass matrix $M$;
            number of HMC iterations for training $N_1$;
            number of samples to generate after training $N_2$;
            probability of generating proposal from VI distribution $p_{\mathrm {jump}}$
            }}
\State{$i = 0$}
\While{$i \leq N_1$}  \Comment Phase 1, Learning
    \State{$\bz_{i+1} \leftarrow \mathrm{HMC\, step}(\bz_i, \pi, \epsilon, L, H, M)$}
    \State{Sample batch $\mathcal{B}=\{\bz_{(1)}...\bz_{(B)}\}$ uniformly from $\{\bz_{1}...\bz_{i}\}$} 
    \State{$\mathcal{L} = -\sum_B\,\log q(\bz_{(i)}; \bnu)$}
    \State{$\bnu \leftarrow \bnu - \epsilon \nabla_{\bnu} \mathcal{L}$}  \Comment{Optimization}
\EndWhile

\While{$i \leq N_1+N_2$}  \Comment Phase 2, Sampling
    \If{$\mathrm{Uniform}(0, 1) \geq p_{\mathrm {jump}}$}
        \State{$\bz_{i+1} \leftarrow \mathrm{HMC\, step}(\bz_i, \pi, \epsilon, L, H, M)$}
    \Else
        \State{$\bz \sim  \log q(\bz; \bnu)$}
        \State{$\alpha =  \frac{\pi^*(\bz) q(\bz_i; \bnu)}{\pi^*(\bz_i) q(\bz; \bnu)}$}
        \State{$\bz_{i+1} \leftarrow \bz$ with probability $\alpha$, otherwise $\bz_{i+1} \leftarrow \bz_i$}
    \EndIf
    \State{Sample batch $\mathcal{B}=\{\bz_{(1)}...\bz_{(B)}\}$ uniformly from $\{\bz_{1}...\bz_{i}\}$} 
    \State{$\mathcal{L} = -\sum_B\, \log q(\bz_{(i)}; \bnu)$}
    \State{$\bnu \leftarrow \bnu - \epsilon_q \nabla_{\bnu} \mathcal{L}$}  \Comment{Optimization}
\EndWhile
\Ensure{$\{\bz_{1}...\bz_{M+N}\},\, q(\bz; \bnu^*)$}
\end{algorithmic}
\end{algorithm}

\section{Variational self-Boosted Sampling (VBS)}
\label{sec:main}

In the previous section, we considered two approaches to infer posterior distribution---
i) HMC which is accurate but prohibitively expensive and 
ii) VI which is fast but inaccurate when using backward KL divergence. 
In this section we propose a hybrid approach which combines VI with sampling \citep{Gabrie21, Naesseth20, Hoffman19}: 
we use the samples generated from HMC to train a variational approximation $q(\bz; \bnu)$ to the target distribution on the fly and in turn simultaneously make proposals from the variational distribution to break correlations in successive samples generated in MCMC \citep{Gabrie21}. 
As the variational approximation improves over iterations, it will also become a good proposal kernel for the Monte Carlo chain and its samples will be readily accepted, thus reducing the correlation length of HMC.
We call this scheme variational self-boosted sampling (VBS).

\subsection{Algorithm}
Starting from a sample point in the target distribution, our algorithm can broadly be divided into two phases\footnote{Here we have assumed that we begin with an initial sample from the target distribution.
If instead we do not have such a sample, there is a warmup or burn-in phase to initialize from a random point and reach such a sample from the target distribution. However since this is identical to HMC, we do not include it explicitly as a part of the algorithm.}- i) learning phase and ii) sampling phase. The full algorithm is presented in Algorithm \ref{alg:hybrid} but briefly, the two phases are:

\begin{itemize}
    \item \textbf{Phase I, Learning Phase}:\\
    In this phase we only run vanilla HMC chains to generate samples from the exact posterior while simultaneously using these samples to learn the variational parameters using Eq.~\ref{eq:ikl*}.
    We do not thin HMC samples i.e. we use all the samples which can be correlated.
    This phase lasts until the variational approximation learns the distribution of the current samples.
    The computational cost of this phase is practically the same as HMC since the training cost is sub-dominant. 
    \item \textbf{Phase II, Hybrid Sampling}:\\
    In this phase we alternate between (with some pre-chosen probability, $p_{\mathrm{jump}}$) making proposals from HMC kernel and the variational distribution.
    We need a criterion to accept these proposals to ensure that we sample from the correct target distribution. 
    For HMC proposals, this is satisfied with the same detailed balance condition as before, Eq.~\ref{eq:DB}.
    For variational proposals, we discuss this criterion in detail below. 
    At the same time, we continue to update the variational distribution with both, the new and the old samples from the learning phase.
    This phase lasts until we have the requisite number of independent samples.
\end{itemize}

Note that since we continuously adapt our variational distribution,
our approach is not strictly Markovian.
However if the adaptation decreases with iterations, 
our approach also becomes Markovian asymptotically.
Then for our algorithm to enjoy the asymptotic correctness of MCMC algorithms,
we need a detailed balance condition for the acceptance of variational proposals\footnote{We assume here that the learnt variational distribution is ergodic, which is the second required condition}, 
similar to HMC.

Let $\bz_1$ be the current sample and $\bz_2$ be the proposal from the variational distribution $q(\bz, \bnu)$.
Then the detailed balance condition is met if the acceptance probability $\alpha$ of making the transition $\bz_1 \rightarrow \bz_2$ is 
\begin{equation}
    \alpha = \mathrm{min}\Bigg(1, \frac{\pi^*(\bz_2)q(\bz_1;\bnu)}{\pi^*(\bz_1)q(\bz_2;\bnu)} \Bigg)
\label{eq:DB_NF}
\end{equation}
This is the balance condition to correctly sample from the distribution $\pi^*(\bz)$, which should ideally be the target unnormalized posterior probability of the sample, $\pi(\bz, \by_0)$. 

However we find in our experiments that due to the high dimensional nature of our posterior distribution and incomplete exploration of this distribution in the learning phase with the correlated samples, 
the variance in acceptance probability is quite high and makes the scheme inefficient. To reduce this variance, we re-scale the target unnormalized posterior probability with the number of grid points $\pi^*(\bz) = \pi(\bz, \by_0)^{1/N^3}$.
In this view, we then consider the learnt variational distribution to be \emph{a proposal distribution for MCMC wherein we can quickly reach samples from the target by running a short chain starting from this proposed point}.
This is why we alternate between variational proposal and HMC proposal with a pre-set probability $p_{\mathrm{jump}}$.
We find that $p_{\mathrm{jump}} \sim 0.2$ gives a good balance between the quality of samples and the acceptance rate of proposals generated from the variational distribution.

\begin{figure}[t]
\begin{center}
\resizebox{\columnwidth}{!}{\includegraphics{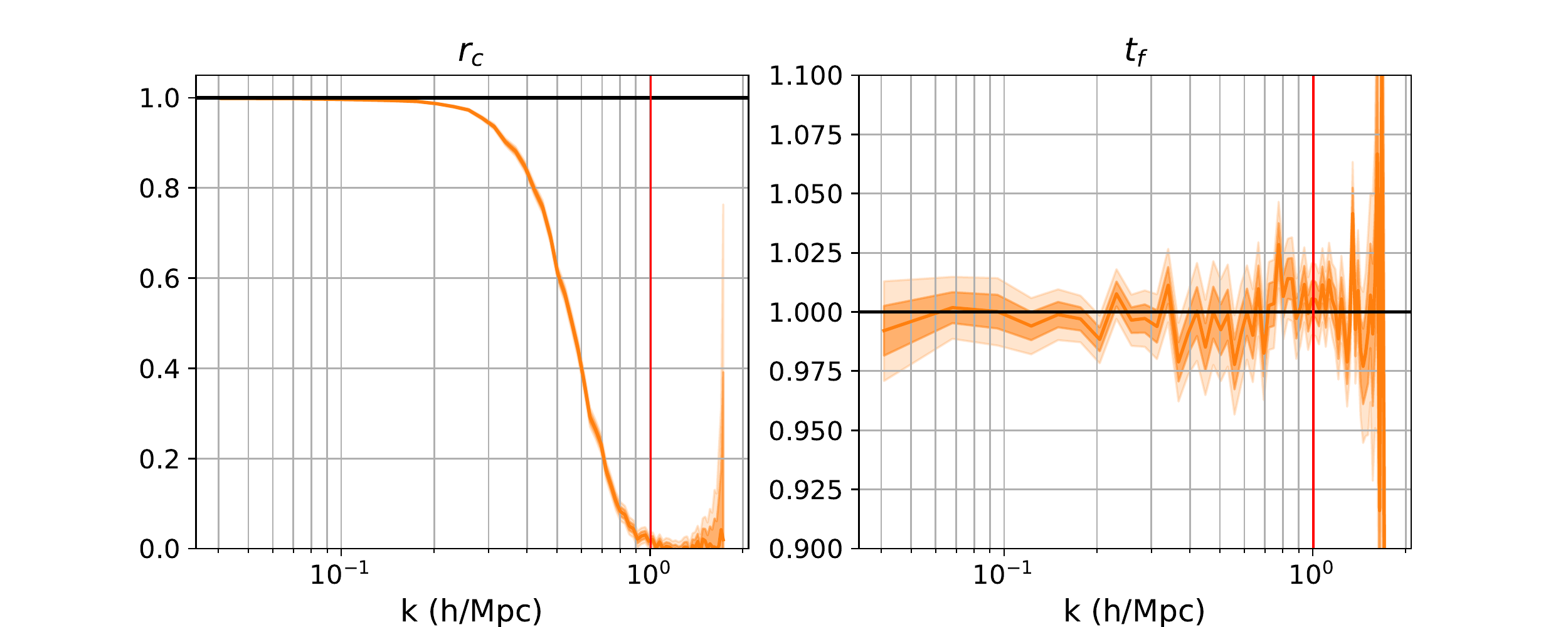}}
\end{center}
\caption{Posterior explored with VBS: We show the cross correlation and transfer function for 1500 consecutive samples from Phase II of hybrid sampling. The results are shown for our fiducial configuration of L=200 Mpc/h and N=64 simulation. The vertical red line represents nyquist frequency.}
\label{fig:hybrid}
\end{figure}

\section{Results}
\label{sec:results}

\begin{figure}[h]
\begin{center}
\hspace*{-1cm}\includegraphics[scale=0.6]{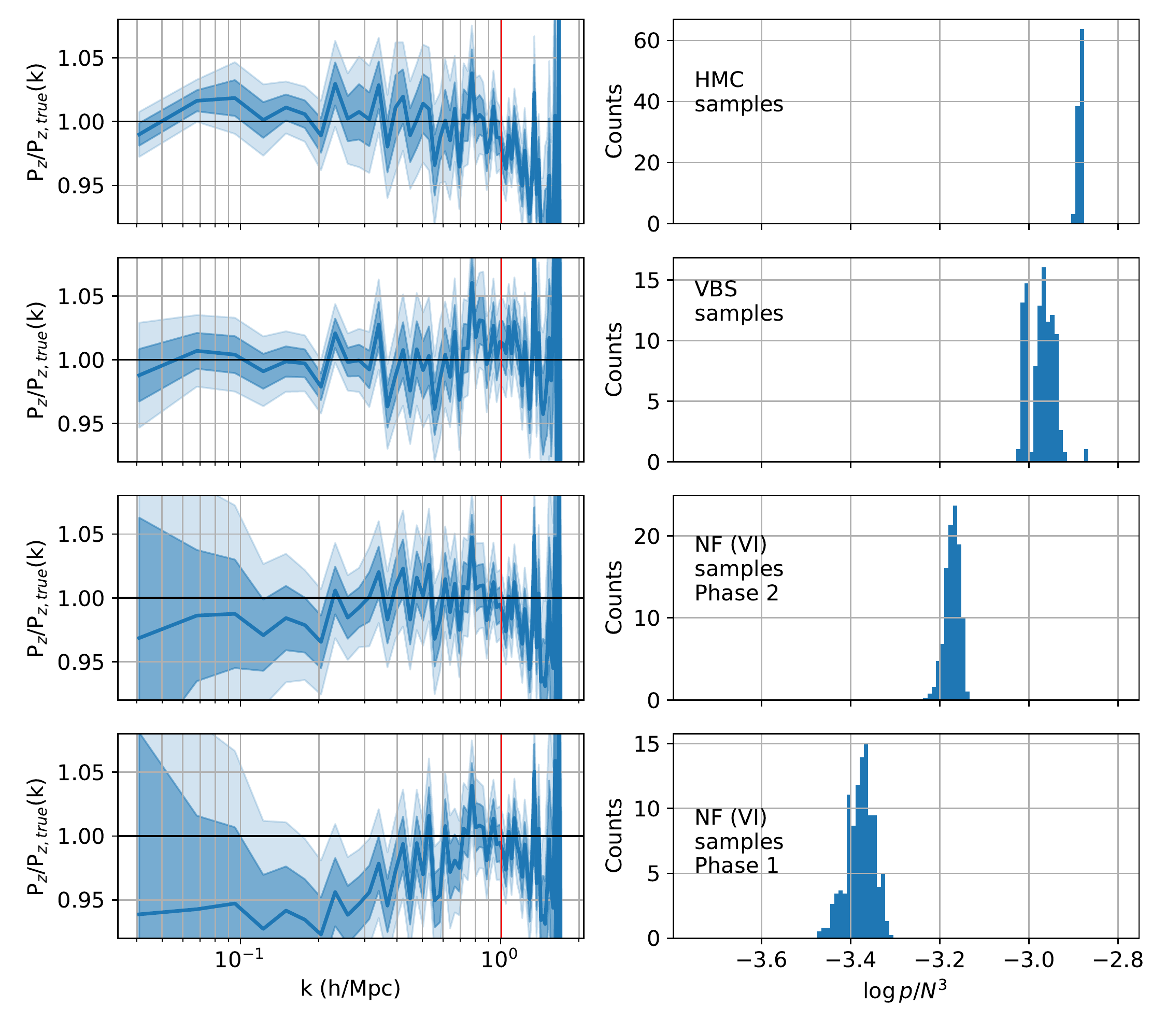}
\end{center}
\caption{Comparing the posterior with different approaches: (left) We show the mean (solids), one- and two-standard deviation (shaded regions) of transfer function for samples of the phase field ($\bz$) from different approaches for our fiducial configuration of L=200 Mpc/h and N=64.
The vertical red line represents nyquist frequency.
(right) We show the distribution of log posterior probabilities, $\log\, p$, for the samples generated by different approaches.
(first row) Vanilla HMC where we have used 800 samples (4 chains, 200 samples each after thinning by factor 20). This acts as the benchmark.
(second row) VBS samples where we use 4000 samples (4 chains, 1000 samples each without thinning).
(third row) Samples generated by the variational distribution (NF) at the end of the second phase. 
(fourth row)  Samples generated by the variational distribution (NF) at the end of the first phase.
This corresponds to a variational distribution fit with forward KL loss, Eq.~\ref{eq:ikl*}.
Though both VBS samples and VI samples after the second phase have transfer function consistent with unity, 
VBS samples are of higher quality since the distribution of their $\log\, p$ is much closer to the benchmark HMC samples.
}
\label{fig:hybrid-compare}
\end{figure}

In this section, we present results for our proposed VBS scheme.
Based on the discussion of merits and issues for HMC and VI in section \ref{sec:hmc} and \ref{sec:vi} respectively, as well as our motivations for the VBS scheme, our goal here will be three-fold:
i) to verify that the samples generated by our scheme are correct and
follow the same distribution as HMC samples, 
ii) to establish that the learnt variational distribution with forward KL is still insufficient and is improved upon with short HMC chains, and finally 
iii) to gauge the gains of our hybrid scheme over HMC as measured with the auto-correlation length of the chains.

\subsection{Verifying VBS posterior}

We begin with verifying the posterior of the hybrid approach. 
As for HMC, we run 4 chains in parallel.
However unlike HMC, these chains are coupled since the samples from all the chains are used to train the NF at each update step and then proposals from the NF are generated for each chain in the sampling phase. 
Based on our experiments, we set $p_{\mathrm{jump}}=0.2$ and the number of samples in training phase $N_1$=500.
Note that this value of $N_1$ leads to only 2 independent samples or less on the largest scales (see Figure \ref{fig:hmc-ac}) and hence we are initially training the NF with mostly correlated samples.
We find the performance of our scheme to be quite robust to $M$ and $p_{\mathrm{jump}}$ within reasonable limits.
We use the same parameters as before for the HMC step of the hybrid scheme i.e.~$\eps$ fit by dual averaging scheme \citep{NUTS} and the number of leapfrog steps $L$ is chosen uniformly between 25 and 50 for every proposal. 

Figure \ref{fig:hybrid} shows the distribution of the summary statistics for the samples generated with VBS scheme
for the baseline case we have been considering, L=200 Mpc/h and N=64 simulations.
We find that both the transfer function and the cross correlation match the expected behavior with the former being unity on all scales and the latter only on the signal dominated scales. 
Moreover, the scatter in the transfer function across scales is similar to that of HMC samples, pointing towards the two distributions being consistent. 

\subsection{Importance of short MCMC chain}

Our next goal is to show that the hybrid nature of our scheme is indeed important
and simply using HMC samples to train the variational distribution with a forward KL loss 
as proposed in Section \ref{sec:fvi} is insufficient. 

To this end, we take the NF trained at the end of phase 1 and phase 2 and generate samples from them. 
In Figure \ref{fig:hybrid-compare}, we compare the distribution of the transfer function (left column) of these samples
with the HMC and VBS samples.
On the largest scales, the samples generated from either NF show a much larger variance than either of the sampling schemes. 

In the right panel, we also show distribution of the unnormalized $\log\, p$ values, i.e. the target posterior probabilities of these samples. 
The samples from NF after phase 1 (learning phase) are of the poorest quality since the training samples generated so far are correlated and explore only a tiny region in the distribution.
Samples generated from NF after phase 2 are better since now the NF has been trained over more independent samples from a larger region in the distribution.
However the VBS samples are still markedly better than the either and much closer to the HMC samples.
HMC samples do still have slightly higher probability than VBS samples because our DB criterion (Eq.~\ref{eq:DB_NF}) of accepting NF proposals in phase II is based on a re-scaled version of the target distribution and not the exact target itself.

This shows that while it is not completely accurate to interpret the variational distribution as having learnt the target distribution,
it can still serve as a good proposal distribution and the short HMC chains do improve the quality of our inference. 

\subsection{Cost of VBS}

\begin{figure}[h]
\begin{center}
\begin{subfigure}{\textwidth}
{\hspace*{-0.6cm}\includegraphics[scale=0.43]{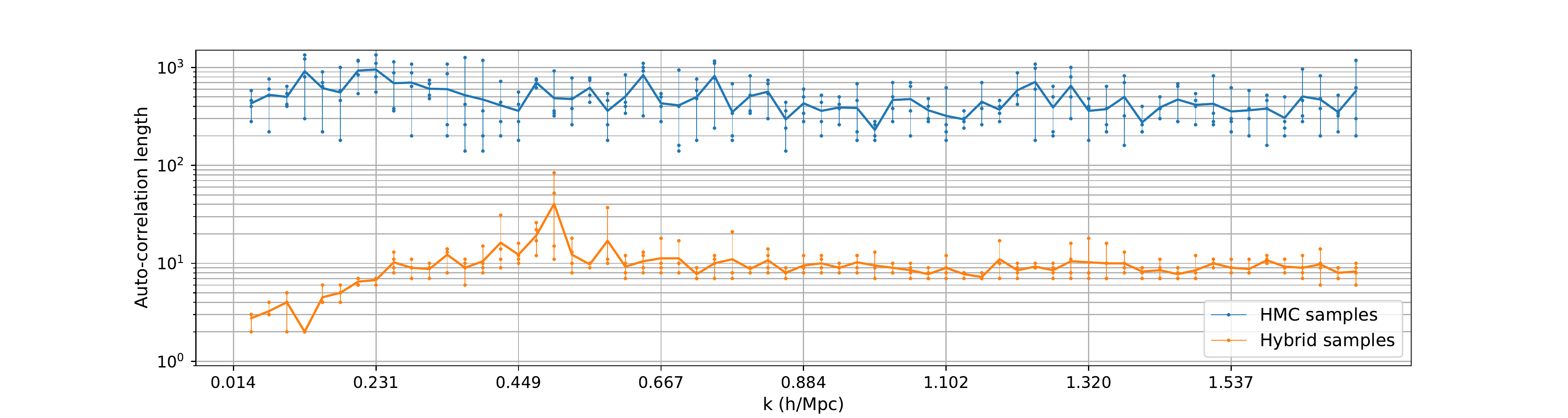}
\caption{High signal-to-noise case (fiducial example): L=200 Mpc/h, N=64}
}
\end{subfigure}
\newline
\begin{subfigure}{\textwidth}
{\hspace*{-0.6cm}\includegraphics[scale=0.43]{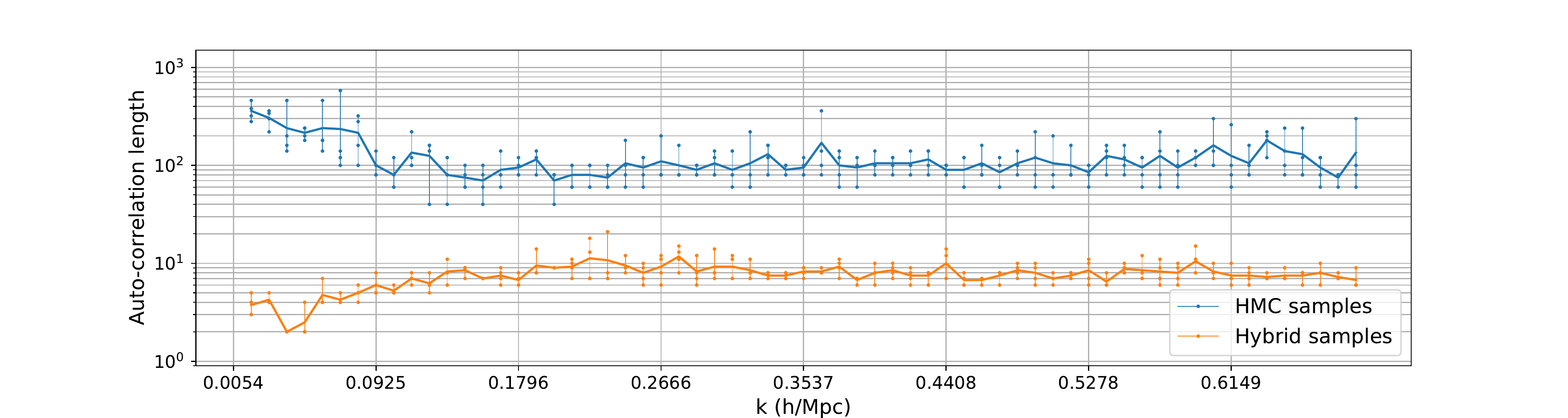}
\caption{Low signal-to-noise case: L=500 Mpc/h, N=64}
}
\end{subfigure}
\newline
\begin{subfigure}{\textwidth}
{\hspace*{-0.6cm}\includegraphics[scale=0.43]{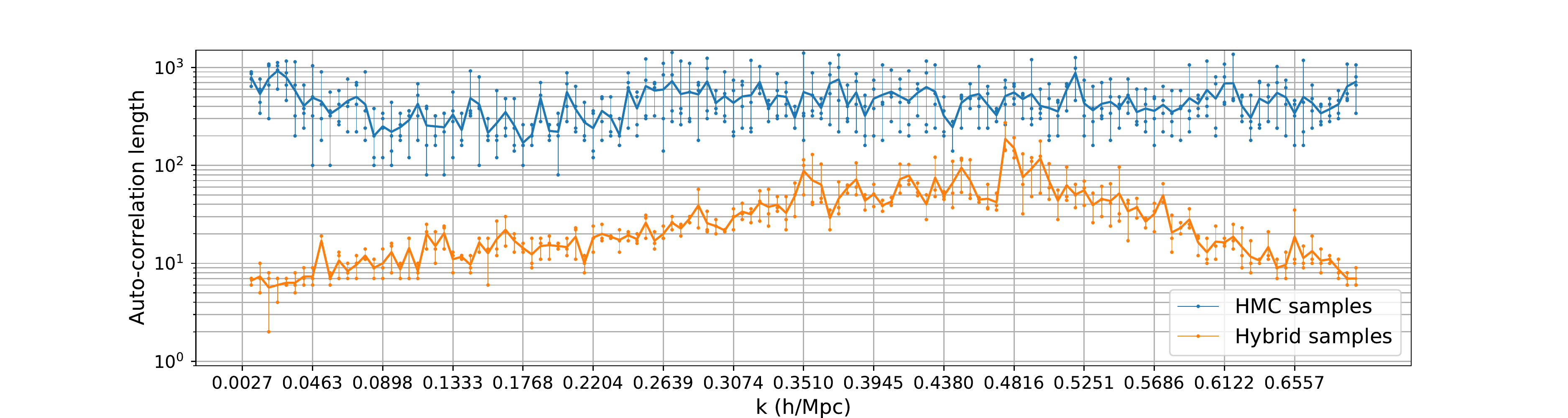}
\caption{Scaling up the low signal-to-noise case: L=1000 Mpc/h, N=128}
}
\end{subfigure}
\end{center}
\caption{Auto-correlation length for different experiments (panels)
as measured with the power in different modes for HMC chains and hybrid samples. 
Different points along the same vertical (k-mode) are four different chains.
}
\label{fig:hybrid-ac}
\end{figure}

Having established the correctness of the samples generated by VBS
and having shown that they are of higher quality than VI,
we next compare VBS with HMC in terms of their efficacy in Figure \ref{fig:hybrid-ac}.
Here we show the auto-correlation length of samples for following configurations of the box size (L) and the mesh (N):
(L, N) = (200 Mpc/h, 64), (500 Mpc/h, 64), and (1000 Mpc/h, 128).
The first two of these experiments have different shot noise levels that allow us to compare the effect of signal-to-noise ratio (SNR) in our data.
The second and the third case have the same shot noise but allow us to see how well our approach scales to larger problems (N). 

For N=64, the auto-correlation length of HMC samples is order hundreds for high SNR case and order tens for low SNR case.
This is consistent with the expectation that the posterior distribution is more complex in high signal regime and hence harder to sample. 
On the other hand, the auto-correlation length of hybrid samples is order tens in both the cases. 
Also note that while the average gains are 40x for low noise case and 10x for high noise case, 
they are larger on the largest scales which are the most correlated for HMC samples.

Finally, as expected, N=128 case is more challenging than N=64 case for both the algorithms.
However VBS still gains a factor of at least 5-50x over HMC depending on the scale under consideration. 
The auto-correlation length in hybrid scheme also shows an interesting feature of increasing until the scale where the SNR$\sim$1 and then dropping again. 
In the future work, we will investigate how this affects the inference of cosmological parameters.

\section{Conclusions}
\label{sec:conclusion}

Forward modeling approaches face the challenging task of doing inference in high dimensions where we need to marginalize over millions of latent parameters which are the phases of the Gaussian initial conditions.
Hamiltonian Monte Carlo (HMC) approaches generate correct samples from the posterior distribution but are prohibitively expensive due to long auto-correlation lengths in the sample chains (Figure \ref{fig:hmc-ac}). 
On the other hand, variational inference (VI) is fast but the backward KL loss can sometimes not be constraining enough to learn the correct target distribution (Figure \ref{fig:bbvi}). 
In this work, we build upon recent works in statistics and machine learning literature to develop variational self-boosted sampling (VBS) --- a hybrid scheme that combines HMC with VI to reap the benefits of both the approaches. 
Thus our approach can be seen as learning the proposal kernel for HMC or alternatively as a variational approximation to the target distribution with short chains serving to correct the learnt approximation. 

We parameterize our variational distribution with a normalizing flow (NF)
which has a learnable mean field Gaussian as the base distribution
and transport layers of alternating Fourier convolutions and element-wise operations. We show that this architecture is flexible enough to learn a low-dimensional distribution of the HMC samples from the posterior which have correct cross-correlation and transfer function (Figure \ref{fig:nflearn}), making it a promising proposal kernel. 

We run experiments on a toy model with the dark matter density corrupted with Gaussian noise as the data.
We run three configurations of the box size (L) and the mesh (N), with (L, N) = (200 Mpc/h, 64), (500 Mpc/h, 64), and (1000 Mpc/h, 128), to explore the effect of the noise in the data as well as the scaling of our normalizing flow and hybrid scheme.
We verify that the transfer function and cross correlation of VBS samples follow a similar distribution to the HMC samples.
We also show that the VBS samples are of a higher quality than the trained NF samples (trained with forward KL) thus demonstrating the benefits of short MCMC chains. 
As compared HMC, we find that in the sampling phase, the auto-correlation length is reduced by a factor of $\sim$50 for the low noise case and factor of $\sim 10$ for high noise case, primarily due to the longer auto-correlation lengths of HMC in the former.
At the same time, both the schemes find it more challenging to N=128 but the hybrid scheme still sees gains of order tens over HMC (Figure \ref{fig:hybrid-ac}). 

In this work we have fixed the cosmology parameters to true cosmology and focused only on inferring the posterior of the phases of the initial conditions since that provides the major challenge of high dimensional inference. 
In the follow-up works, we plan to extend the hybrid sampling to sample the cosmology parameters and the phases simultaneously. 
We have also used simple forward models such as ZA for computational ease, and dark matter density as the data observable but we expect our conclusions to qualitatively remain the same for more realistic gravity models as well as tracers such as halos and weak lensing maps.

\subsubsection*{Acknowledgments}
CM would like to thank Francois Lanusse for the initial discussions at the conception of the project and help with VI code, Marylou Gabrie for sharing insights on hybrid sampling approach, Biwei Dai for sharing his TRENF code. CM would also like to thank Uros Seljak, Vanessa Boehm, Ben Wandelt, Shirley Ho, Kaze Wong, Wenda Zhou for useful discussions and Vanessa Boehm again for detailed comments on the manuscript.

\bibliography{biblio}
\bibliographystyle{JHEP}

\appendix

\section{Normalizing Flow}
\label{app:nflows}

Normalizing flows transform a simple base distribution $q_B$
with a transport map $T_{\theta}$ consisting of a series of invertible, bijective mappings into more complex distributions of interest \citep{Kobyzev20}. 
We use NF to parameterize our variational family such that it is flexible enough to capture the target distribution
\begin{equation}
    q(\bz; \bnu) = q_B(T_\theta^{-1}(\bz); \bnu_B) |\det \nabla_{\bz} T^{-1}_\theta |
\end{equation}
where the parameters of the base distribution and the transport map compose our variational parameters $\bnu=\{\bnu_B, \theta\}$.

\subsection{Base Distribution}
Traditionally when NF are used to learn generative models,
the base distribution consists of a simple distribution with few or no trainable parameters, such as a standard normal.
However in our case, the target is the posterior of a specific data realization and this breaks the symmetry of the target distribution.
Hence for our base distribution, we choose the mean-field normal i.e. $q(\bz; \bnu_B) := \mathcal{N}(\bmu, \bm{\Sigma})$ where $\bmu$ and $\bm{\Sigma}$ are now the same shape and size as the phase field, i.e., N$^3$ grids. 
In our experiments, we find that while fixing $\bm{\Sigma}=1$ does not affect our posterior accuracy significantly, 
however keeping $\bmu$ trainable is crucial for any meaningful inference.

\subsection{Transport Map}
The transport map consists of a series of invertible transformations such that the log-determinant of their Jacobain can be estimated quickly.
Hence NF typically use specialized coupling layers or autoregressive layers \citep{Dinh16, papamakarios17}.
However these NF scale poorly to three dimensional data and large (millions) parameter spaces. 

We take an alternate approach for our transport map that was recently shown to accurately learn the high dimensional data likelihood of cosmological fields in \cite{TRENF}. 
Motivated by the fact that the cosmological fields are rotationally and translationally invariant, \cite{TRENF} propose constructing transport maps using Fourier-space convolutions. 

\subsubsection{Fourier Space Convolutions}
A convolution in configuration space can be performed as a product with a transfer function $t(\bk)$ in the Fourier space. 
This transfer function can be element-wise and hence of the same dimensionality N$^3$ as the parameters. 
However for rotational and translational invariant fields, 
the transfer function becomes only a function of scales, $t(k)$,
which can be parameterized by a few tens of parameters. 
Moreover since the transformation consists of simply multiplying be a scalar function, the Jacobian is straightforward to estimate.

Thus the overall transformation for a configuration space field $\bx$ is
\begin{equation}
    \bx' = \mathcal{F}^{-1} (t_{\theta}(k) \mathcal{F}(\bx))
\end{equation}
where $\theta$ are the learnable (variational) parameters and $\mathcal{F}$ is the Fourier transform operation.
The transfer function can be any interpolation function and we model it as a Cubic Hermite polynomial.
Then, the knots values and slopes at knot positions constitute $\theta$.

\subsubsection{Element-wise Transformations}
We alternate the Fourier space convolutions with learnable element-wise transformations $\Psi_{\phi}$ in the configuration space. 

The simplest such transformations are affine (scale and shift) transformations
\begin{equation}
    \bx' = \alpha\bx + \beta
\end{equation}
with $\phi = \{\alpha, \beta\}$ as the scale and shift variational parameters.
We consider two cases- 
i) global affine transformations wherein $\alpha$ and $\beta$ are scalars and the entire field is shifted and scaled uniformly, or 
ii) mean-field affine transformations wherein $\boldsymbol{\alpha}$ and $\boldsymbol{\beta}$ are now $N^3$ grids, same as the parameters $\bx$. 
While ii) increases the number of parameters of our NF by a lot, it allows us break the constraint of rotational and translational invariances in our transport map that is made by using Fourier space convolutions.
We find that for N=64, global affine transformations sufficed but for N=128 using mean-field affine transformations markedly improved the quality of inference. 

Affine transformations are linear but the element-wise transformations can also be made non-linear. For instance, \cite{TRENF} used monotonic rational-quadratic splines as non-linear transformations.
However in our experiments, using splines instead of linear transformations did not seem to significantly affect the quality of posteriors for our toy model and hence we did not use them for the current experiments.

\subsection{Learnt Distribution}
Every layer of our NF consists of a Fourier space convolution followed by an element-wise operation to construct a unit transformation $f=\bx_0 \rightarrow \bx_1$:
\begin{equation}
    \bx_1 = \Psi_{\phi}(\mathcal{F}^{-1} (t_{\theta_1}(k) \mathcal{F}(\bx_0)))
\end{equation} 
These layers can be stacked and are combined with the base distribution to parameterize our target distribution. 

\begin{figure}[h]
\begin{center}
\includegraphics[scale=0.7]{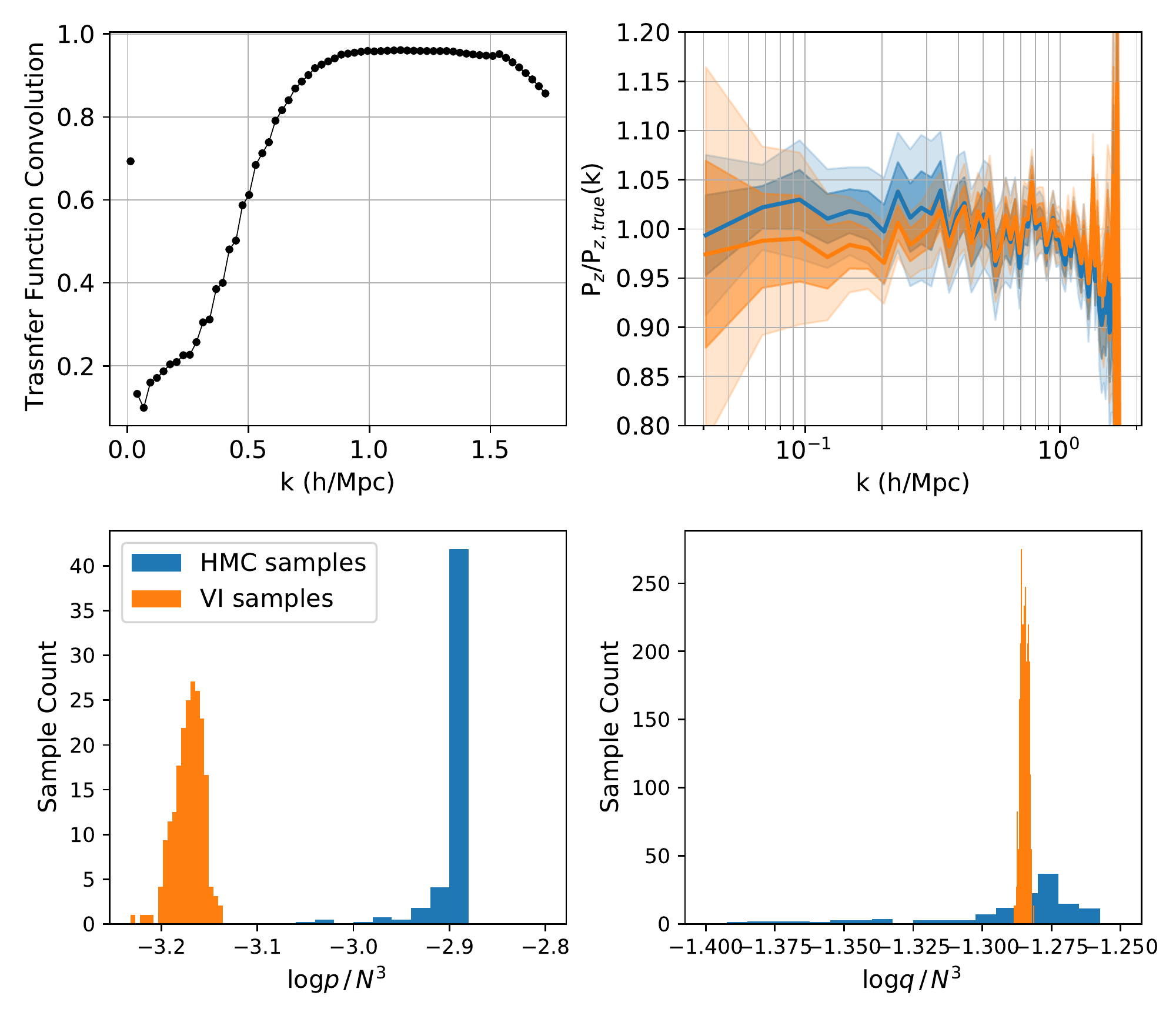}
\end{center}
\caption{Learning of normalizing flow: (top left) The transfer function learnt by the Fourier convolution layer of NF. 
(top right) Distribution of the transfer function for samples from HMC on which NF is trained (blue) and the samples generated by the normalizing flow (orange). 
(bottom left) Distribution of unnormalized $\log p$, the true posterior probability of HMC samples (blue) and samples from the trained NF (orange).
(bottom right) Distribution of $\log q$, the variational posterior probability as estimated by NF of HMC samples (blue) and samples from the trained NF (orange). 
}
\label{fig:nflearn}
\end{figure}

It is crucial for variational inference to ensure that the parametric family is flexible enough to capture the target distribution.
To verify this, we train our NF on the independent samples generated from HMC. 
and show different metrics to gauge its performance in Figure \ref{fig:nflearn}. We show the results for L=200 Mpc/h box and N=64 grid. 
The first panel shows the transfer spectra learnt by the Fourier convolution layer and it seems that the convolution layer acts as a high pass filter. 
The second panel shows the distribution of the power spectra of HMC samples and samples generated from the trained normalizing flow. 
The NF samples have higher variance than HMC samples but are still consistent with unity on all scales. 
Note that this is similar to Figure \ref{fig:hybrid-compare} except that the NF there is trained on VBS samples while in this case it is trained on HMC samples. 
While not shown here, we have verified that the distribution of the cross-correlation of the NF samples with the true initial conditions is similar to that of HMC samples. 
This, combined with the visual consistency of all samples, can lead one to conclude that NF learns the high dimensional target distribution. 

In the bottom panel of Figure \ref{fig:nflearn}, 
we examine the learnt NF in greater detail.
The two panels show the histogram of 
the unnormalized logarithm of probability under the target distribution (posterior, $\log\, p$) 
and the learnt variational or NF distribution ($\log\, q$) for HMC and NF samples\footnote{Note that since these probabilities are not normalized by the evidence, these should not be compared directly across panels}.
It's immediately apparent that the two distributions are different and in more ways than can be explained by simple normalization.
This, coupled with the fact that the posterior distribution of the 
transfer function and cross-correlations for the two sample sets is somewhat consistent leads us to conclude that while the NF does not learn the full high dimensional target distribution, it does correctly learn a lower-dimensional manifold. 
Hence instead of using NF to learn the posterior of the phase fields completely, 
we instead use the variational distribution as a proposal kernel in the hybrid scheme where short HMC chains propagate the generated samples to the target distribution. 

\end{document}

%% file: math_commands.tex
\usepackage{amsmath,amsfonts,bm}

\def\eqref#1{equation~\ref{#1}}

\def\1{\bm{1}}

\def\eps{{\epsilon}}

\DeclareMathAlphabet{\mathsfit}{\encodingdefault}{\sfdefault}{m}{sl}
\SetMathAlphabet{\mathsfit}{bold}{\encodingdefault}{\sfdefault}{bx}{n}

